\documentclass[a4paper]{article}
\usepackage{amssymb,amsmath,amsthm} 
\usepackage{mathrsfs}
\usepackage{graphicx}
\usepackage{times}
\usepackage{mathtools}
\usepackage{tikz}
\usepackage{blindtext}
\usepackage{hyperref}
\usepackage{siunitx}
\usepackage{tabularx}
\usepackage{multirow}
\usetikzlibrary{shapes.geometric, arrows}
\usepackage{xcolor}
\pagecolor{white}
\graphicspath{ {./images/} }
\usepackage{caption}
\usepackage{subcaption}
\usepackage{cases}
\usepackage{appendix}

\usepackage{authblk}
\usepackage{arxiv}
\usepackage{indentfirst}
\usepackage{url}
\usepackage{orcidlink}

\begin{document}

\title{A seven-step guide to spatial, agent-based modelling of tumour evolution}

\date{}

\author[1]{Blair Colyer\,\orcidlink{0000-0002-9940-5707}}
\author[1]{Maciej Bak\,\orcidlink{0000-0003-1361-7301}}
\author[2]{David Basanta\,\orcidlink{0000-0002-8527-0776}}
\author[1]{Robert Noble*\,\orcidlink{0000-0002-8057-4252}}
\affil[1]{Department of Mathematics, City, University of London, London, UK}
\affil[2]{Department of Integrated Mathematical Oncology, H. Lee Moffitt Cancer Center and Research Institute, \protect\\ Tampa, Florida, USA}
\affil[*]{robert.noble@city.ac.uk} 

\maketitle


\begin{abstract}
{Spatial agent-based models are increasingly used to investigate the evolution of solid tumours subject to localised cell-cell interactions and microenvironmental heterogeneity. Here we present a non-technical step by step guide to developing such a model from first principles, aimed at both aspiring modellers and other biologists and oncologists who wish to understand the assumptions and limitations of this approach. Stressing the importance of tailoring the model structure to that of the biological system, we describe methods of increasing complexity, from the basic Eden growth model up to off-lattice simulations with diffusible factors. We examine choices that unavoidably arise in model design, such as implementation, parameterisation, visualisation, and reproducibility. Each topic is illustrated with examples drawn from recent research studies and state of the art modelling platforms. We emphasise the benefits of simpler models that aim to match the complexity of the phenomena of interest, rather than that of the entire biological system.}
\end{abstract}


\section*{Introduction}

Cancer initiation, progression, and treatment responses are Darwinian evolutionary processes \cite{casas2011cancer, merlo2006cancer} that can be investigated using a wide range of mathematical and computational methods. Examples include evolutionary game theory \cite{yang2016nonlinear, basanta2011role}, branching processes \cite{danesh2012branching, durrett2010evolutionary}, and Moran processes \cite{west2016evolutionary, durrett2016spatial}. Yet while many tools have yielded important insights into cancer evolution, the study of spatial aspects -- especially important in carcinomas, constituting the majority of humans cancers -- often necessitates a spatially explicit approach, such as a spatial agent-based model.

An agent-based (or individual-based) model is a computational model of a system made up of autonomous, interacting ``agents". Spatial agent-based models (SABMs) have long been used to study the evolution of spatially structured communities because they can reveal how the processes of selection, drift, and gene flow depend on localised interactions among agents (typically individual organisms) or between agents and their spatially varying environment. As new technologies generate better spatial tumour data, SABMs are proving ever more useful in oncology. Typical applications include understanding tumour development, inferring the effects of driver mutations, and predicting treatment outcomes. For example in recent studies, Aif {\it et al.} \cite{aif2022evolutionary} used an SABM to investigate the evolutionary rescue of drug-resistant tumour subclones; Saha {\it et al.} \cite{saha2023silico} used an SABM to investigate adaptive cancer therapy; and Bull and Byrne \cite{bull2023quantification} used an SABM to simulate interactions between macrophages and tumour cells.

To support this burgeoning research field, here we present a seven-step guide to designing and implementing spatial agent-based models in which the agents are locally-interacting tumour cells or cell subpopulations. Starting from the simplest cellular automata, we discuss options for adding greater complexity and biological realism, such as multi-level spatial structure and environmental heterogeneity. Based on our extensive experience of developing and using SABMs \cite{noble2022spatial, noble2020and, bak2023warlock, bacevic2017spatial}, we cover practical issues such as event scheduling, visualisation, and how to use SABMs to infer parameter values from experimental or clinical data. Each topic is illustrated with examples from our own demon-warlock modelling framework \cite{bak2023warlock, noble2022spatial}, other state of the art modelling platforms, and studies that have used SABMs in cancer research. Whereas our focus is on tumour evolution, much of our advice applies equally to similar modelling methods used to study bacterial colonies, invasive species, and organismal development. The guide is designed to be accessible for biologists and clinicians without specialist mathematical knowledge.

\section{Spatial structure}

Spatial structure determines the evolutionary balance between selection and drift, the nature of gene flow between subpopulations, and the strength of ecological interactions. When a model fails to accurately represent the spatial structure of a biological system, the model's predictions and inferences for that system may be highly unreliable \cite{noble2022spatial, strobl2022spatial}. It follows that the parameters of spatial structure -- such as the size of locally interacting cell communities and the manner of cell dispersal -- should be accorded the same importance as evolutionary parameters in model design. Notwithstanding the trade-off between model simplicity and realism, spatial structure parameters should, as far as possible, be derived or inferred from empirical data.
 
\subsection{Stochastic cellular automata}

\begin{figure}
\centering
\includegraphics[scale=0.75]{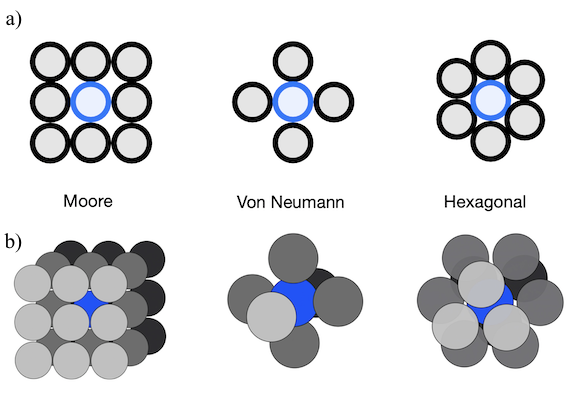}
\caption{Some common neighbourhoods that govern the update rules for cellular automata and other agent-based models in two dimensions ({\bf a}) and three dimensions ({\bf b}). A focal agent (cell) is shown in blue and its neighbourhood sites in grey.}
\label{neigh}
\end{figure}

Many of the simplest spatial agent-based models are cellular automata. A cellular automaton is a model that plays out on a grid of sites in one or more dimensions. Each site is associated with one of a set of at least two possible states. Each site also belongs to a subset of sites called a neighbourhood, of which some examples are shown in Figure~\ref{neigh}. For example, the von Neumann neighbourhood in two dimensions contains the nearest sites in the cardinal directions (up, down, left and right). A cellular automaton sequentially updates itself according to a set of rules. The update rules for a given site depend on its own current state and the states of the sites in its neighbourhood.

Whereas the update rules of many cellular automata are deterministic \cite{schiff2011cellular}, probabilistic rules are more appropriate for modelling stochastic processes such as biological evolution. Because its next state depends only on its current state, a stochastic cellular automaton is equivalent to a collection of locally interacting Markov chains.

In biological terms, each state corresponds a type of cancer cell or some other entity (such an immune cell or part of the extracelluar matrix). Generally we will assume that the focal agents in our models are cancer cells and we will use the terms ``agent" and ``cell" interchangeably where appropriate. A cellular automaton permits a cell's event probabilities (for example, its division, death, and dispersal rates) to depend on the number of neighbouring cells, thus accounting for crowding or Allee effects. Event rates can also vary according to the types of the neighbouring cells, for example to simulate cell competition or immune predation.

Models of asynchronous processes, such as cell division in a tumour, typically use asynchronous updating, meaning that only one or a small number of sites are modified per update \cite{louis2018probabilistic}. In addition to being more realistic, asynchronous updating is often necessary to prevent conflicts. For instance, if two cells are attempting to divide but only one space is available for the two potential daughter cells then one must take priority.

\subsection{The Eden growth model}


Among the simplest stochastic cellular automata is the Eden growth model. This model is typically implemented on a two- or three-dimensional regular square grid with only two possible states: unoccupied ($S_0$) and occupied ($S_1$). With each iteration, the update rule causes a site in the neighbourhood of an $S_1$ site to switch from $S_0$ to $S_1$. In this way new $S_1$ sites (cells) are added to the surface of a cluster. The Eden growth model on an $n$-dimensional grid self-organises to resemble an $n$-dimensional ball with a non-trivial surface. The growth curve of the $S_1$ population approaches a polynomial of degree $n$ \cite{eden1961two}.

The three most popular options for the Eden growth model update rule can be labelled alphabetically:
\begin{itemize}
\item {\bf A}vailable site-focussed: Choose at random an $S_0$ site in the neighbourhood of an $S_1$ site, and switch it from $S_0$ to $S_1$. 
\item {\bf B}ond-focussed: Choose at random an $S_1$ site with a probability proportional to the number of $S_0$ sites in its neighbourhood, and then randomly choose an $S_0$ neighbour and switch it to $S_1$. 
\item {\bf C}ell-focussed: Choose at random an $S_1$ site with at least one $S_0$ site in its neighbourhood, and then randomly choose an $S_0$ neighbour and switch it to $S_1$. 
\end{itemize}

Although these update rules result in similar large-scale patterns, they generate cluster surfaces with different microscopic properties. Indentations in the model surface are more likely to be filled, and spikes are less likely to form, under option C than under option B, and under option B than under option A. Hence option C generates the smoothest surface and option A the roughest \cite{jullien1985scaling}.

Variants of the Eden growth model have been used to investigate the evolution of paediatric glioma \cite{tari2022quantification}, colon cancer \cite{hamis2021targeting}, hepatocellular carcinoma \cite{waclaw2015spatial}, and non-small cell lung cancer \cite{jagiella2016inferring}. Many studies use a variant that includes stochastic cell death. By opening up spaces for cell division, cell deaths increase clonal mixing within the tumour and facilitate selection \cite{waclaw2015spatial}.

\subsection{Other grid-based stochastic cellular automata}

Other stochastic cellular automata can be more appropriate than the Eden growth model for modelling systems in which state changes are not confined to the surface. Spatial branching processes are similar to Eden growth models except that if a dividing cell has no space to divide then it can create space by budging other cells. An intermediate model can be created by stipulating that only nearby cells can be budged, so as to simulate physical constraints on cell division. Chkhaidze \textit{et al.} \cite{chkhaidze2019spatially} recently used such a model to investigate how spatially constrained tumour growth alters signatures of clonal selection and genetic drift in cancer genomic data. Good practice is to implement budging along an approximately straight line between the dividing cell and the nearest empty site. If budging is instead restricted to the cardinal directions or the cardinal and intercardinal directions then the simulated tumour will self-organise into an approximate square or octahedron, rather than a more biologically plausible disc or ball.

Another option is to allow dividing cells to replace, rather than displace, their neighbours. In the voter model, the update rule is such that, with a certain probability, a randomly selected site copies the state of a neighbouring site. Biasses can be introduced by setting unequal copying probabilities, corresponding to differences in cell fitness. Simple (linear) voter models satisfy a convenient property called coalescing duality, which means that their typical behaviour can be explained through mathematical analysis \cite{durrett2007random}. In a pioneering 1972 study, Williams and Bjerknes \cite{williams1972stochastic} used a biassed voter model to simulate the spread of skin cancer through the basal epithelial layer.

The cellular Potts model (CPM), also known as the Glazier-Graner-Hogeweg model \cite{graner1992simulation, savill1997modelling}, more explicitly simulates physical interactions among cells and between cells and their microenvironment. The model takes place on a lattice and each cell is represented by multiple lattice sites (as opposed to only one lattice site, as in previously discussed models), corresponding to the cell's volume. Cells are deformable and can adhere to one another or to surrounding empty sites (which might represent extracellular matrix or growth medium). Hamiltonian mechanics describe the overall energy of the system depending on adhesion forces and resistance to changes in cell volume. A random lattice site is chosen at each time step and its state is copied to a random neighbouring site. If the new configuration has lower energy than the previous configuration then the change is always accepted; otherwise, the probability of accepting the change depends on the Boltzmann temperature. The CPM has been used in numerous cancer studies, such as for simulating tumour growth, invasion and evolution \cite{szabo2013cellular}, or for investigating how cell compressibility, motility and contact inhibition shape tumour cell clusters \cite{li2014effects}. The CompuCell3D modelling environment {compucell} provides an efficient, flexible CPM implementation.

The biological lattice gas cellular automaton \cite{deutsch2005mathematical} excels instead at modelling cellular movement, and especially collective migration, in a simple, computationally efficient, and physically correct fashion. The model must play out on a square or hexagonal lattice in 2 dimensions, or a cubic, dodecahedral or icosahedral lattice in 3 dimensions. States incorporate cell velocities. For instance, consider a 2-dimensional square lattice in which each site contains 5 nodes: one for each directional velocity and a resting node at the centre. A cell occupying any one of these nodes can divide into other nodes on the same site. A cell can also reorient itself by moving between nodes on the same site, and can move between sites according to its velocity, provided there is space to do so. This model has been used, for example, to give insights into breast cancer invasion plasticity \cite{deutsch2021bio}.

\subsection{Multi-level spatial structures}

An important limitation of all the aforementioned cellular automata is that their uniform spatial structures are inconsistent with the biology of many tumour types. Various common cancers have glandular structures and grow via individual cells or small cell clusters invading neighbouring tissue \cite{pandya2017modes, lugli2021tumour}. Colorectal adenomas are also glandular but grow through gland fission \cite{preston2003bottom}.

Inspired by classical population genetics models \cite{moran1958random}, a simple, conventional way to account for multi-level spatial structure in tumours is to assign cells to local subpopulations, called demes, located on a regular grid. Thus each grid site is allowed to contain not only one but dozens, hundreds, or thousands of cells. The subpopulation size per deme is prevented from exceeding a certain threshold -- known as the deme's carrying capacity -- by decreasing cell division rates or increasing death rates as the subpopulation size grows.

Deme-based models allow for more complicated modes of cell dispersal. As in the voter model, cells can be assigned some probability of invading neighbouring demes, either individually or in clusters. The dispersal probability can also be made to depend on the population of the deme being invaded, so that cells disperse more easily in less densely populated regions near the tumour periphery. Alternatively, each occupied deme can be assigned a probability of undergoing fission, resulting in some of its cells being moved to an unoccupied neighbouring deme. Depending on the degree of budging allowed, the deme-level dynamics of the fission model can resemble an Eden growth model (no budging of demes) or a spatial branching process (unlimited budging). Deme-based models additionally allow for the explicit simulation of tissue invasion, such that a tumour can grow only via its cells invading demes that are initially filled with normal cells \cite{noble2022spatial}.

\subsection{Aggregating agents}

If the within-deme subpopulations can be assumed to be well-mixed then cells that belong to the same deme and have the same phenotype and genotype can be modelled collectively, rather than as individual agents. This model design not only improves computational efficiency but can also facilitate mathematical analysis. For example, when cells disperse by invading neighbouring demes, the dynamics of the demon-warlock framework are approximately equivalent to the well understood spatial Moran process. Cells can be randomly selected within a deme by sampling from a hypergeometric distribution.

Even greater efficiency can be realised by not modelling inter-deme dynamics at all, and simply making the demes themselves the model agents \cite{sottoriva2015big, siegmund2009inferring}. Although such coarse-graining enables the simulation of much larger tumours, it comes at the cost of reduced precision. Care should be taken in translating between mutation rates per cell and effective mutation rates per deme.

\subsection{Off-lattice models}
Instead of confining agents to a regular grid, we might instead locate them in continuous space. This structure is potentially more realistic but also entails more parameters, more decisions to be made, and typically higher computational costs \cite{beerenwinkel2015cancer}. To prevent multiple cells occupying the same space and to maintain tumour integrity, we now must model the movement of cells in response to physical forces such as cellular adhesion and repulsion \cite{franz2014travelling}. We may also choose to model directed movement under the influence of diffusible factors (hapotaxis).

There are several practical ways to prevent cells overlapping in an off-lattice model, depending on how the agents are implemented. Suppose we have spherical cells, each with fixed radius $r$. We can then specify that when, as a result of cell division or movement, the distance between two cells' centres is less than $2r$, both cells will simply be pushed in opposite directions. Alternatively, to account for cell deformation, we might implement repulsion only when the distance between cell centres falls below some threshold value smaller than $2r$ \cite{macklin2012patient}. Some modelling platforms achieve greater realism and tractability by implementing adhesion and repulsion forces using functions rooted in physics, which are beyond the scope of this guide (see documentation cited in the appendix).

\section{Mutation}

Having chosen an appropriate spatial structure, we next will decide which cell phenotypes and genotypes to include in our state space, and how to model mutations between these states. As ever, the goal is to balance model simplicity, realism, and computational demands.

\subsection{Defining phenotypes}

A good part of the difficulty in designing a useful model stems from the fact that much of the experimental data gathered by cancer biologists focusses on genetic mutations while the rules that govern the behaviour of the agents in an SABM assume an understanding of the key cancer phenotypes. The most basic actions a tumour cell might perform at any given time step are apoptosis/death, proliferation, and motility. These are often considered as simple probabilistic events and often modelled in a exclusionary manner, so that if a cell is moving then it is neither proliferating nor dying. The required probabilities can either be taken directly from experimental data (which is often hard to measure \textit{in vivo} and unrealistic \textit{in vitro}) or calibrated with \textit{in vivo} pre-clinical models. 

Using hard-coded rules to model the phenotype of a tumor cell, while relatively simple, does not capture the flexibility shown by biological cells in the mapping between genotype and phenotype. Gerlee and colleagues have instead proposed capturing some of the complexity of this mapping by embedding neural networks inside each agent, so that the phenotype emerges in a non-linear way as a result of the agent's state and the different microenvironmental inputs to which the agent is receptive \cite{gerlee2009modelling}.
\subsection{Trait evolution versus population (epi)genetic models}

Once phenotypes have been defined, the next step is to determine how these phenotypes will change as a result of mutations. One option is to model mutations as phenotypic switches. Many studies consider models with only two possible tumour cell states -- mutated and unmutated -- which differ in fitness \cite{sottoriva2015big}, degree of drug resistance \cite{gallaher2018spatial}, or some other trait. Grow-or-go models assume that cells can reversibly switch between predominantly migratory and predominantly proliferative phenotypes \cite{hoek2008vivo}. Other models examine the evolution of continuous traits, such as levels of glycolysis and acid production \cite{robertson2015impact}.

If we are more interested in clonal dynamics then we can explicitly track changes to the (epi)genome. These mutations are conventionally assigned to three groups according to how they affect cell fitness: driver mutations (which increase cell fitness), passenger mutations (no effect), and deleterious mutations (negative effect). For simplicity, most studies assume an infinite sites model \cite{kimura1969number}, such that no two mutations can occur at the same site. Finite sites models must be parameterised based on observed mutation frequencies \cite{schenck2022homeostasis}.

\subsection{Example: The Eden growth model with mutation}

\begin{figure}
\centering
\includegraphics[scale=0.25]{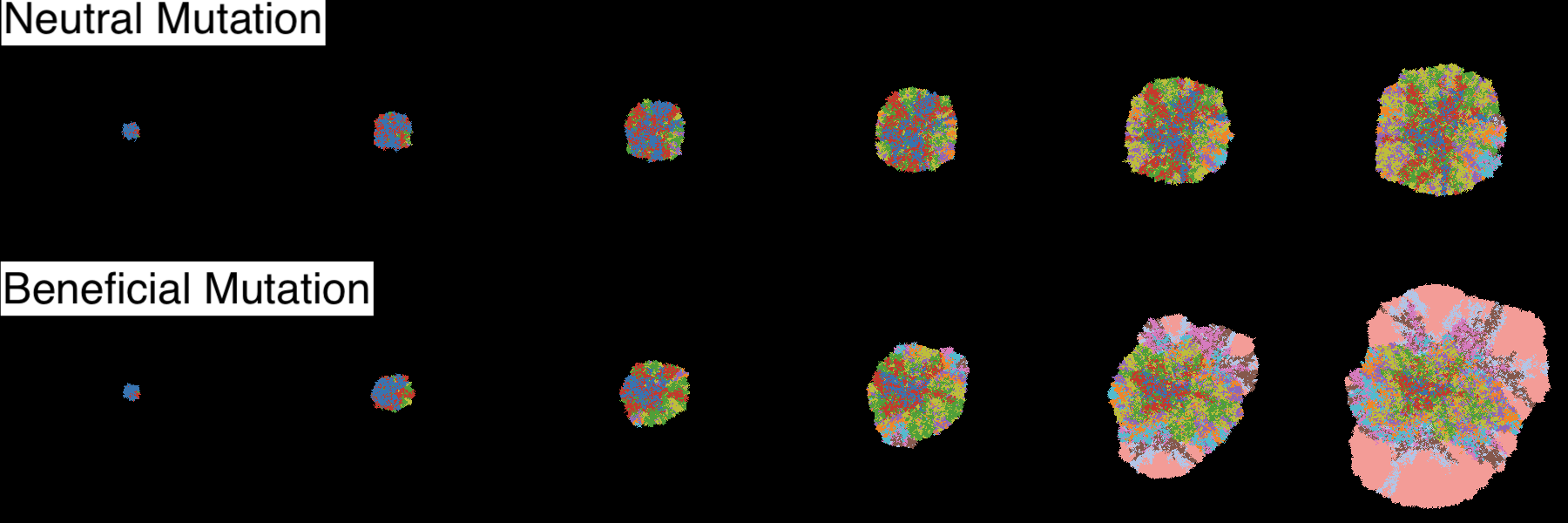}
\caption{The result of running an Eden growth model with nearly neutral mutations (top) and beneficial mutations (bottom). Model produced in HAL using some in-built examples as a skeleton for the code. \cite{westhal}.}
\label{haleden}
\end{figure}

We can convert an Eden model into an evolutionary model by implementing mutation. The grid and neighbourhood are defined as before but now we have multiple cell states 
$S_1, S_2, S_3, \dots$ and mutation rates between each pair of distinct cell states. A simple option, assuming infinite sites, is to set all mutation rates to be zero except in the case of $S_i$ to $S_i+1$ for all $i \ge 0$, so that every $S_i$ cell has exactly $i$ mutations. Let us assume that all these mutations are drivers and their effects combine multiplicatively, such that each mutation increases the division rate by a factor of $1+s$, with $s \ge 0$. Assume also that mutation occurs only at the time of cell division, and the number of new mutations per daughter cell is Poisson distributed. We then arrive at a reasonable toy model of spatial tumour evolution that can be implemented in not much more than 100 lines of code, as we illustrate with an R script \cite{noblegithub}. Figure~\ref{haleden} shows results of implementing a similar model in the HAL platform \cite{westhal}.

\subsection{Distributions of fitness effects}

Modelling the evolution of a quantitative trait, such as cell division or death rate, leads to further design decisions. As in our toy model, it can be wiser to draw mutation fitness effects from a probability distribution instead of setting them all equal. To see why, consider a model of an expanding tumour that, in the absence of mutation, has radius growth rate $c_0$, and in which the spread of mutants is not confined to the periphery (for example, a biassed voter model). When a fitter mutant arises within the wildtype population, its long-term fate, in the absence of further mutation, will be sensitive to its radius growth rate, $c_1$. If $c_1 < c_0$ then the mutant will remain forever rare; if $c_1 > c_0$ then the mutant is bound to take over the entire tumour; if $c_1 = c_0$ then the mutant will become relatively more abundant over time without ever fully replacing the wildtype. Randomising the fitness effect randomises $c_1$ and so randomises mutant fates. The demon-warlock framework draws each selection coefficient (relative increase in cell division rate) from an exponential distribution.

Strictly multiplicative fitness is best avoided in all but the smallest-scale models as it can lead to unrealistically high fitness values. This is especially problematic if mutation is implemented at the point of cell division, which creates a feedback loop in which lineage fitness grows at an ever increasing rate. A simple solution implemented in the demon-warlock framework is diminishing returns epistasis. When the selection coefficient of a driver mutation is $s$, instead of multiplying the division rate by $1+s$, we instead multiply by $1+s(1-b/b_{max})$, where $b$ is the previous division rate and $b_{max}$ is an upper bound.

\section{Event scheduling}

The next step is to consider how to implement cell events algorithmically. Event scheduling can be the most important factor in determining computational efficiency, especially in simpler grid-based models. The optimal choice strikes a balance between efficiency, simplicity, and biological realism.

\subsection{Gillespie's algorithm}

The Gillespie Stochastic Simulation Algorithm \cite{gillespie1976general} is an especially simple and popular solution to event scheduling. Event rates are assumed to depend only on the current state of the model and the time between events is exponentially distributed (as in a Poisson process), such that two events cannot occur simultaneously. The steps of the algorithm are as follows: 

\begin{enumerate}

\item Initialise the system. 
\item Set event rates (birth rates, death rates, dispersal rates, etc.).
\item Randomly determine the next event such that $\mathbb{P}(event = E) = rate(E) \big/ \Sigma (rates)$
\item Implement the chosen event.
\item Advance the timer by $\delta t \sim \textrm{Exp}(1 \big/ \Sigma (rates))$
\item Repeat from step 2 until a stop condition is reached.

\end{enumerate}

This algorithm is more efficient than the event timer approach (see below) and is very easy to implement. In statistical terms, the simulated sequence of events corresponds to a trajectory of a set of stochastic differential equations, called the master equations. This means we have a good mathematical understanding of how the algorithm behaves.

Our toy Eden growth model \cite{noblegithub} provides an example implementation of Gillespie's algorithm. This model further improves computational efficiency by keeping track of the cells that have space to divide, so that the next dividing cell can be chosen from among this subset (which in $n$ dimensions scales with the radius to the power of $n-1$) rather than from the entire cell population (which scales with the radius to the power of $n$). The drawback is that cells without space to divide never undergo mutation, which may be an unjustifiable assumption in a serious research model.

Modifications of Gillespie's algorithm, such as tau leaping \cite{gillespie2001approximate}, are even faster but less accurate. Tau leaping allows multiple events to occur simultaneously, which may be problematic in a spatial model if the events affect multiple sites in close proximity (for example, if two cells are chosen to divide into the same empty site). Moreover, tau leaping improves performance only when the system is dominated by a small number of large, homogeneous subpopulations, which is typically not the case in SABMs.

\subsection{Gillespie's algorithm with phase-type distributions}

A shortcoming of the Gillespie algorithm is that some events, such as cell division, are not true Poisson processes with exponentially distributed waiting times. In effect, the Gillespie algorithm permits arbitrarily short cell cycles. Some cells may divide several times while, in the same period, others with identical division rates fail to divide at all.

One way to achieve more realistic cell cycle periods without sacrificing very much computational efficiency is to use a phase-type probability distribution, such as an Erlang distribution, constructed using a mixture of exponential distributions. This entails executing the Gillespie algorithm as above except that when a cell is selected for division it doesn't necessarily divide immediately but instead changes its position in the cell cycle. Given a target probability distribution for cell cycle periods, we can use an algorithm to choose transition rates such that the resulting phase-type distribution has the same mean, variance, and skew as the target \cite{osogami2006closed}. For example, suppose that all cells begin in division state 0. When a cell is selected (according to a state-dependent probability), its state is updated. When a state 3 cell is selected it divides and both progeny are reset to state 0 \cite{belluccini2022counting}. The method's greater realism comes at the cost of additional memory demands and longer execution time, compared to the basic Gillespie algorithm.

\subsection{Random sampling with binary trees}

\begin{figure}
\centering
\includegraphics[scale=0.45]{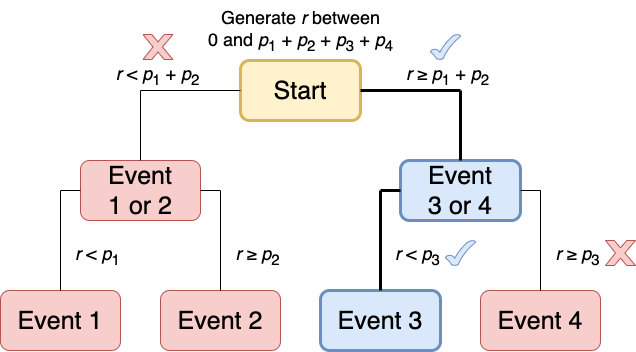}
\caption{An example of using a binary tree to select an event (Event 3) from four options. Selected nodes are shown in blue.}
\label{binary}
\end{figure}

When we have more than a handful of events to choose from it will be much more efficient to implement event selection using a binary tree. Suppose, for example, that we have four possible events with rates $p_1, p_2, p_3$ and $p_4$. If we store the rate sums $p_1 + p_2$, $p_3 + p_4$, and $p_1 + p_2 + p_3 + p_4$ then we can choose an event as follows. First we generate a random number $r$ from a uniform distribution between $0$ and $p_1+p_2+p_3+p_4$, and we examine whether $r < p_1+p_2$. Supposing $r$ is greater than $p_1+p_2$, we then test whether it is less than $p_3$. If so then we choose event 3; otherwise event 4. Effectively, we have traversed a binary tree, beginning at the root node associated with the sum of all event rates, and ending at a terminal node associated with a single event (Figure~\ref{binary}).

The binary tree method is efficient because both the number of steps needed to choose an event, and the number of nodes that need updating following a change in an event rate, grow only with the logarithm of the number of possible events. For example, we need only twenty steps to choose between one million possible events. As long as the cell population keeps growing, there is little benefit to pruning nodes and it is easy to ensure that the tree remains balanced. The rate sums together take up only as much computer memory as the individual rates. The main costs are in terms of code development time and code complexity. Binary trees require careful implementation and error checking to ensure that existing nodes are updated and, when required, new nodes are added after each model event. The demon model implements binary trees and periodically recalculates event rate sums to prevent excessive accumulation of rounding errors.

\subsection{Cell cycle timers} 

A less efficient alternative to using phase-type distributions is to draw inter-division times directly from a chosen probability distribution. This approach enables more precise tracking and adjustment of individual cell cycles. An algorithm used in recent studies \cite{robertson2015impact, gallaher2018spatial} is as follows:

\begin{enumerate}

\item Initially assign every cell $i$ a countdown timer set to time $t_i$ drawn from some probability distribution (dependent on the  cell's phenotype).
\item Subtract $\delta t$ from every countdown timer, where $\delta t \ll t_i$ for all $i$.
\item For all cells $i$, in random order: \\
	 A. Implement cell death and dispersal events for $i$;\\
	 B. If $i$ is alive, has space to divide, and $t_i \leq 0$, then $i$ divides; \\
	 C. Assign each new cell a countdown timer, set to some random time dependent on the new cell's phenotype.
\item Repeat from step two until a stop condition is reached. 

\end{enumerate}


How much this approach reduces computational efficiency will depend on other aspects of the model. It is likely to be much slower than a well implemented Gillespie algorithm when applied to a simple grid-based model, due to the additional burdens of updating every cell (Step 2) and shuffling all the cells (Step 3) at each small time step. In an off-lattice model, where cells move much more frequently than they divide, and where a shuffling algorithm may already be required to randomise the order in which cell positions are updated, the cost of updating cell division state at the same time as position may be negligible.


\section{Microenvironment}

Whereas many SABM studies focus on the effects of spatial structure and cell-cell interactions, real tumours evolve in a complex microenvironment that varies over space and time. 
This tumour microenvironment, comprising both molecular elements, such as cytokines, and other (non-cancer) cells, constitutes the cancer ecosystem \cite{anderson2020tumor} -- a key element of the selection process driving somatic evolution.
Given a good rationale and sufficient parameterisaton data, we may choose to extend our model by explicitly simulating microenvironmental factors in the form of agents (in the case of immune cells or stromal cells) or diffusible factors (such as oxygen and drugs). Permitting cancer cells to modify their selective environment creates potential for emergent complexity and niche construction \cite{chaplain1996mathematical, qi1993cellular}.

\subsection{Hybrid cellular automata}

Hybrid cellular automata (or HCA) have been used to model interactions between tumour cells and diffusible factors for more than twenty years. As described in a pioneering 2001 paper by Patel and colleagues \cite{patel2001cellular}, these models consist of two interdependent components: stochastic cell events, and deterministic reaction-diffusion partial differential equations. The latter component dictates how chemicals or other factors work their way through the system as they are consumed and processed by cells. Local concentrations of diffusible factors contribute to the cell update rules.

Typically we assume that diffusible factor concentrations rapidly re-equilibriate following changes in the configuration of cells. We can then numerically solve the equations to find the equilibrium concentrations either after every cell event or, trading some accuracy for greater efficiency, after a relatively small number of cell events have occurred. Suitable procedures for solving partial differential equations as initial value problems can readily be found in textbooks and software libraries. These range from simple but inefficient algorithms based on the classical Gauss-Seidel method, which require only a few dozen lines of code \cite{press2007numerical, patel2001cellular, bacevic2017spatial}, to the highly sophisticated BioFVM solver \cite{ghaffarizadeh2016biofvm}, which is specifically optimised for hybrid SABMs. Several SABM platforms include their own methods for solving reaction-diffusion equations in two or three dimensions (see appendix).

\subsection{Types of diffusible factor}

To add biological realism, we might make cell division and death rates in our model depend on the local oxygen and glucose concentrations as these factors diffuse through the tumour from the surrounding medium (in very small tumours and tumour spheroids) or from point sources representing blood vessels (in larger, vascularised tumours). We might also modify dispersal rates so that cells follow oxygen or glucose gradients. Potential adverse factors include acid produced through tumour cell metabolism, and drugs that diffuse from blood vessels. Hybrid cellular automata are especially suitable when the supply of an influential factor is highly variable over space or time, such as in the case of intermittent drug treatment \cite{bravo2020hybrid}.

\section{Parameterisation and inference}

Although theoretical models can be valuable for generating hypotheses and providing proof of concept, if we want to apply an SABM to studying a particular biological system then we must ensure that its influential parameter values are set appropriately. Parameterisation should ideally be based on clinical or experimental data specific to the biological system of interest; otherwise values can be estimated from studies of similar systems or theoretical considerations (for instance, diffusion coefficients approximately correlate with molecular weight). Influential parameters might pertain to the effects of mutations, drugs, oxygen and glucose; rates of chemical supply, diffusion, consumption and decay; cell dispersal modes and rates; baseline cell death rates, crowding effects and the size of interacting cell communities. Since calibrating SABMs is often computationally demanding, high-performance computation may be required to generate the necessary resources to calibrate them properly. 

\subsection{Example: Hybrid cellular automaton for simulating a tumour spheroid}

Bacevic and Noble {\it et al} \cite{bacevic2017spatial} parameterised a HCA to mimic tumour spheroid evolution under drug treatment. In spheroids the limiting factor for cell survival and proliferation is oxygen. Other diffusible factors such as glucose were therefore omitted to simplify the model without compromising its usefulness. The oxygen concentration in the medium and oxygen diffusion rates were drawn from previous studies \cite{casciari1992mathematical, kim2007hybrid, grimes2014method}, as were the mathematical relationships between oxygen consumption rate, cell proliferation rate and local oxygen concentration \cite{casciari1992variations, grimes2014oxygen}. The different maximum proliferation rates of drug-sensitive and resistant cells, reflecting a fitness cost of resistance, were determined from new monolayer growth assays. Cells with insufficient oxygen supply were assumed to die.

Since oxygen effects alone fail to account for the extent of quiescence observed in tumour spheroids, Bacevic and Noble {\it et al} implemented crowding effects by permitting cell budging only within a specified radius. New monolayer growth assays revealed that the relationships between cell proliferation rate, death rate and drug dose could be well approximated with piecewise linear functions. The drug's impact on proliferation was further assumed to multiply the oxygen effect, consistent with prior observations \cite{casciari1992variations}. Drug consumption was also modelled using Michaelis-Menten kinetics, with a diffusion rate chosen according to the drug's molecular weight and an appropriately low consumption rate. Thus parameterized, the SABM accurately predicted the outcomes of new tumour spheroid experiments \cite{bacevic2017spatial}.

\subsection{Example: Hybrid cellular automaton of the bone ecosystem in cancer}
Araujo and colleagues \cite{araujo2016hybrid} developed a hybrid cellular automaton for which the goal was to capture the ecosystem of the bone. A crude approximation of this ecosystem includes the bone itself, the myeloid-derived cells such as osteoclasts that resorb bone, and the cells derived from messenchymal stem cells, such as osteoblasts, that deposit new bone. Each of these cell types can be modelled as discrete agents regulated by diffusible factors -- such as TGF-$\beta$, RANK ligand, and other factors embedded in the bone matrix -- described by partial differential equations. Parameterisation of the model is facilitated by the fact that non-cancerous cells have more predictable phenotypes, and the model's overall behaviour can be calibrated to ensure it recapitulates bone homeostasis. Araujo and colleagues thus studied how bone metastatic prostate cancer cells could infiltrate the bone ecosystem, take advantage of it, and grow \cite{araujo2018size}. They also investigated what prostate cancer cells in the primary tumour should be of concern to physicians, and why conventional treatments that fail to disrupt tumour-ecosystem interactions also fail to provide long-term cancer cures in bone metastatic prostate cancer \cite{araujo2014integrated}.

\subsection{Parameter inference}

Unknown parameter values can be inferred by combining an SABM with a statistical method. This is, in fact, often the main objective of an SABM study. Approximate Bayesian computation is a popular approach that, in its simplest form, infers the value of a parameter $\theta$ as follows
\begin{enumerate}
\item From our data, calculate some summary statistic $\mu_{data}$;
\item Set $i = 1$;
\item Run the model using a candidate parameter value $\theta_i$ drawn from some prior distribution;
\item Calculate the summary statistic $\mu_i$ for the model output;
\item If the difference between $\mu_i$ and $\mu_{data}$ is less than a predefined tolerance then add $\theta_i$ to the posterior distribution;
\item Increment $i$;
\item If $i$ is less than some threshold then repeat from step 3. 
\end{enumerate}

Although simple in principle, approximate Bayesian computation requires careful implementation. The accuracy and precision of inferences depend on the choices of prior distributions, summary statistics, and tolerances, as well as the number of iterations. Typically multiple parameter values cannot be precisely derived from prior data or models, in which case each should be assigned a vague (high variance) prior distribution. Tolerance values should be tuned such that neither too many nor too few candidate parameter values are accepted to the posterior distribution. Summary statistics should capture features of the system that provide useful information about the parameters of interest. A useful template is a 2010 study \cite{sottoriva2010integrating} in which Sottoriva and Tavar\'{e} inferred aspects of stem cell dynamics in the colonic crypt by combining a cellular Potts model with approximate Bayesian computation, using a summary statistic based on methylation patterns.

An alternative to this approach was recently outlined in \cite{cess2023calibrating}, in which the authors describe a novel method utilising neural networks to reduce both tumour images and SABM simulations to low-dimensional points. The distance between these points acts as a quantitative measure of how the two differ. This enables direct comparison, and by using parameter fitting algorithms to minimise the distance between the two sets of points, parameters can be estimate directly from the images and the simulations. 

\subsection{Sensitivity analysis}

Whatever the objective, an essential step in any modelling study is so examine, as far as is practical, how the results and conclusions depend on uncertain aspects of the model. A common approach is to run a large number of model variants with different combinations of plausible parameter values. Varying one parameter at a time can provide useful insight into which parameters have the greatest impact on model output, with the shortcoming that non-linear interactions between parameters are often neglected. A more sophisticated approach is to infer a multivariable ``metamodel" function -- a model of the model -- that approximately describes how the model's parameters determine its outputs.

Since varying many parameters systematically on a continuous scale is infeasible, sampling methods such as Sobol sequencing \cite{sobol1967distribution} or Latin hypercube sampling \cite{mckay2000comparison} can be used to generate a set of near-randomly sampled combinations of parameter values. Both methods were used in a recent SABM study of cancer cell response to ATR-inhibitors \cite{hamis2021targeting}. A recent introductory review explains specifically how to apply these methods to cancer ABMs \cite{hamis2021uncertainty}. It is important to note that thorough sensitivity analysis involves varying not only parameter values but also mathematical relationships, aspects of spatial structure, and any other influential model components.


\section{Visualisation}

\begin{figure}
\centering
\includegraphics[scale=0.3]{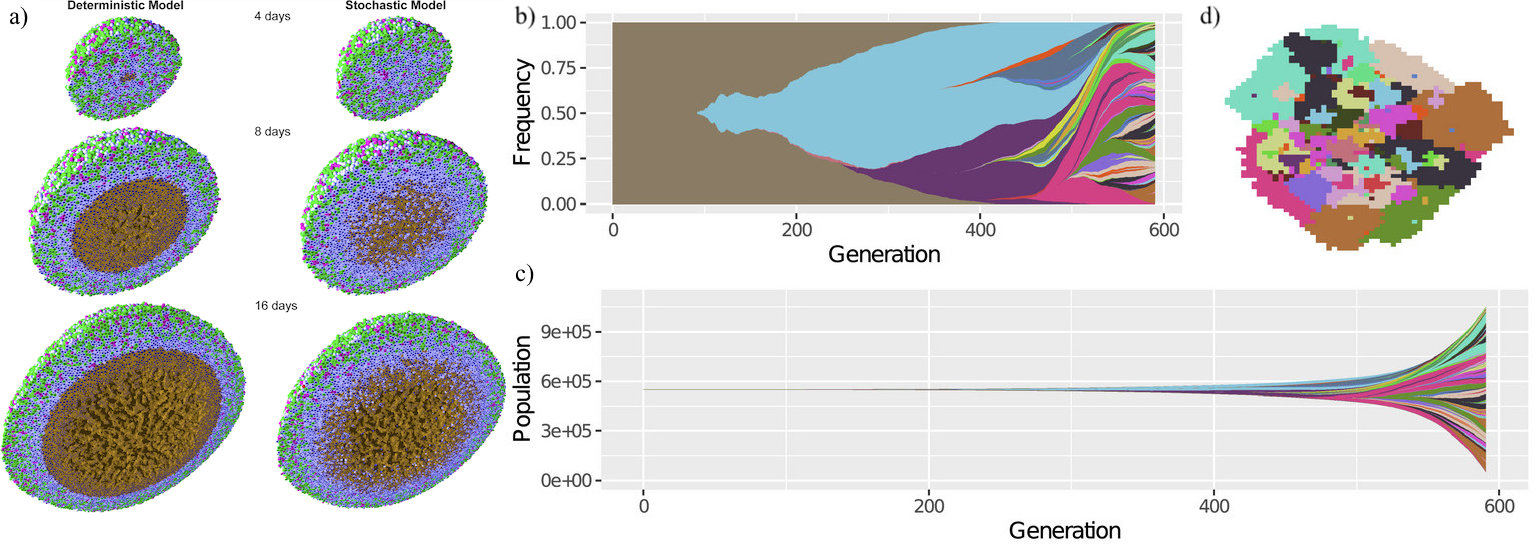}
\caption{{\bf a)} Plots of a 3D off-lattice ABM, produced in PhysiCell \cite{ghaffarizadeh2018physicell}, showing a cross-section of model states of a hanging-drop spheroid growth simulation at different time points, using either a deterministic or a stochastic SABM. Cells are coloured according to cell cycle position. Cells in the $K_1$ cell cycle state are green, post-mitotic $K_2$ cells are magenta, quiescent cells are pale blue, apoptotic cells are red, and necrotic cells are brown. Cell nuclei are shown in dark blue. {\bf b)} Muller plot showing phylogenies and phenotype frequencies over time. {\bf c)} Fish plot showing phylogenies and phenotype population sizes over time. {\bf d)} 2D grid plot corresponding to same simulation as the Muller and fish plots in previous panels, with the same colour scheme, at the final time point. Plots b and c were produced using the R package {\it ggmuller} \cite{ggmuller}. Image a is reproduced from \cite{ghaffarizadeh2018physicell} under the terms of a Creative Commons Attribution License and with the approval of Paul Macklin. Plots b-d are reproduced from \cite{noble2020and} under the terms of a Creative Commons Attribution License.}
\label{visual}
\end{figure}

Having built and parameterised a model, we next require useful ways to visualise its output. Typical methods represent spatial information, multidimensional phenotypic information, or evolutionary dynamics. Representing all these aspects in a single image is generally impossible.

\subsection{Spatial plots}

A spatial plot represents the state of an SABM at a moment in time. Producing a two-dimensional grid plot of a two-dimensional on-lattice model is straightforward -- we simply output the state of each site as a matrix of numbers and input this matrix into a bitmap (or raster) plotting function in R, Python, MATLAB, or similar software, using different colours to represent the different states (Figure~\ref{haleden}). Our toy Eden growth model \cite{noblegithub} provides an example implementation. Diffusible factor concentrations can be shown outside the tumour using a colour gradient and inside the tumour by adjusting brightness \cite{bacevic2017spatial}. We can apply the same method to off-lattice models by specifying a grid and assigning each grid square a value that summarises the states of all points within the square. Given multi-level spatial structure, we can represent the most abundant state in each deme \cite{noble2022spatial}.

Illustrating three-dimensional information is more technically demanding as we need to project the object onto a two-dimensional plane, determine the visible surface, and add shading (as in Figure~\ref{visual}a). Suitable computational methods include rasterisation and ray tracing, which can be performed in R and Python or using dedicated software, such as POV-Ray. Further details can be found in the PhysiCell documentation (see appendix). A much simpler solution is to plot only two-dimensional slices.

\subsection{Visualising evolutionary dynamics}

Muller plots represent subpopulation dynamics and phylogeny, while disregarding spatial information. The horizontal axis represents time and the vertical axis corresponds to subpopulation frequency. Each subpopulation is depicted as a shaded area emerging from its immediate ancestor (Figure~\ref{visual}b). Fish plots are similar but show population size rather than frequency (Figure~\ref{visual}c). Software packages for producing these plots include ggmuller \cite{ggmuller} and EvoFreq \cite{gatenbee2019evofreq}.

\subsection{Phenotype space plots}

In a phenotype space plot, the axes correspond to continuous traits such as cell fitness, metabolic type, and degree of drug resistance, and each point represents a cell. We can visualize phenotypic evolution by animating phenotype space plots from a series of time points. Robertson-Tessi and colleagues pioneered the use of these plots in cancer research in their 2015 study of the effects of metabolic heterogeneity on tumour growth \cite{robertson2015impact}.

\section{Reproducibility}

Reproducibility is a cornerstone of the scientific method. A reproducible modelling study not only allows others to easily regenerate its results but also permits further data processing, downstream analysis of generated data, generation of summary statistics, ease of production for visual representations or plots, and even adaptation of the existing model for novel purposes.

\subsection{Principles of reproducible research} 

Gundersen \cite{gundersen2021fundamental} describes three categories of reproducibility:

\begin{itemize}

\item Outcome reproducibility: The reproduction experiment’s result matches the original. If the same analysis of the result is performed, the same conclusions can be drawn, and the original hypothesis is supported by both experiments.

\item Analysis reproducibility: The reproduction experiment’s result differs from the original, but if the same analysis method is used, the interpretation of the results still matches the original.

\item Interpretation reproducibility: The reproduction experiment’s outcomes and the analysis of said outcomes both differ, but the interpretation matches the original interpretation.

\end{itemize}

Computational modelling studies should typically aim for the highest standard of outcome reproducibility. If care is taken to construct a well-packaged computational study in a controlled digital environment, then in principle, given a suitable machine, the study should be easily reproduced exactly. This entails not only comprehensively explaining methods, results, analyses, and interpretation, but also sharing the model code and scripts used at every step of pre-processing and analysis, providing a detailed description of how to execute the code, and sharing any associated data and parameterisation and configuration files.

In their outline of best practices to observe throughout a computational research project, Sandve and colleagues \cite{sandve2013ten} advocate tracking how every result is produced and reporting intermediate results as well as final outcomes. To make code easier to reproduce, one should catalogue the versions of software used at every point, record the seeds used in any random number generation, and implement version control \cite{heroux2009barely}. Manual data manipulation should be avoided in favour of using automated methods to reformat and process raw data files. The raw data used to produce summary data plots should be easily accessible to facilitate easy plot reproduction and to allow readers to check individual data points. Textual descriptions of methods and results should link to the associated raw data and code so that a reader can easily follow all the steps leading to interpretations. Lastly, modellers are highly encouraged to share each full study, ideally with a dedicated public server. One such research-oriented database is zenodo \cite{zenodourl}, where scientists may freely upload their research output permanently as a citeable piece of software.  

\subsection{Workflow managers, package managers and containers}

A complex computational model will often require multiple steps to be carried out in sequence. If a high-performance computing (HPC) cluster is required to run the model efficiently -- as is typical for complex models -- it is essential to utilise a workflow manager to properly orchestrate the steps \cite{wratten2021reproducible}. Open-source workflow managers allow researchers to package a model into a reproducible, cross-platform workflow. Nextflow \cite{molder2021sustainable} and Snakemake \cite{di2017nextflow} are among the most popular workflow managers with several published pipelines \cite{kieser2020atlas, holzer2021poseidon, zhao2018lncpipe, cornwell2018viper}, strong community support, and extensive documentation, giving users flexibility when designing their own custom pipelines. Snakemake is based on Python, a popular language among computational biologists and bioinformaticians. 
Nextflow uses the Java-based language Groovy, which has a Python-style structure and is relatively easy to for Python users to learn.  Both also enable automatic parallelisation for HPC clusters, which can be essential for complex SABMs or for running multiple instances of smaller models simultaneously. 

Another option is to utilise container technologies, considered by many to be the gold standard in computational research. These are less computationally demanding than running an application on a computer directly or using a virtual machine and so permit faster deployment, patching, and scaling. Containers also allow users to deploy the application on multiple operating systems or machines without reformatting, and will run the application the same way no matter where they are deployed \cite{moreau2023containers}. Docker \cite{merkel2014docker} is a popular container design platform which permits packaging applications into distribution-independent containers. Another option, Bioconda \cite{gruning2018bioconda}, enables easy dependency management, and can be deployed inside a container. 

\subsection{Extendable modelling platforms}

For many research projects, the easiest option can be to build on an existing open-source agent-based modelling platform (see appendix for a brief guide). Some of these platforms -- such as Chaste \cite{mirams2013chaste}, CompuCell3D \cite{swat2012multi}, HAL \cite{westhal} and PhysiCell \cite{ghaffarizadeh2018physicell} -- excel in simulating off-lattice cell populations in complex microenvironments. Others, such as demon \cite{nobledemon} (which has an automated computational workflow, Warlock \cite{bak2023warlock}), J-SPACE \cite{angaroni2022j} and SMITH \cite{streck2023smith}, focus on efficient modelling of evolutionary dynamics. Several are modular platforms, which facilitate reproducibility by making it easy to create and share extensions of the generic software. Nevertheless, even the most flexible platform is necessarily based on certain fundamental assumptions, structures, and algorithms. If we want to create an especially innovative model, requiring several novel components that pre-existing modelling platforms lack, then we might find it best to start from scratch. In principle, specialist rather than generalist models permit greater optimisation in terms of memory demands and execution time.

\subsection{FAIR principles in data management}

As the volume of publicly available research data has been growing exponentially in recent decades \cite{statista2023volume}, proper digital data management and annotation is recognized as an essential step in computational research -- crucial for research reproducibility. Most notably, the FAIR principles have become a cornerstone in modern data management, particularly in the realms of scientific and research data \cite{wilkinson2016fair}. FAIR is an acronym that encapsulates a set of guiding principles: Findable, Accessible, Interoperable, and Reusable. To be FAIR, data must first be Findable, meaning that it is easy for both humans and machines to discover, thanks to comprehensive metadata and proper indexing. Data should be Accessible, ensuring that access rights and permissions are clear and well-defined, thus minimizing barriers to entry. Interoperable data is structured in a way that allows integration with other datasets by adhering to common standards, formats, and vocabularies. Lastly, data should be Reusable, with thorough documentation, contextual information, and availability in a format that facilitates easy replication and reuse. Altogether, the FAIR principles serve as a framework for enhancing data sharing, management, and collaboration, ultimately driving scientific progress and fostering open science initiatives. Major organisations that have embraced FAIR guidelines include the European Open Science Cloud \cite{eoscportal}, the European Life-Science Infrastructure for Biological Information \cite{elixirabout}, the US National Institutes of Health \cite{nihgov}, and the Global Alliance for Genomics and Health \cite{ga4gh}. 

\section*{Discussion}

Having surveyed the numerous choices that arise in any SABM project, we are faced with a problem: how can we choose the most appropriate model? In tumour evolution research, unlike in much of physics and engineering, there is no standard approach. Rather we must tailor a model to each research question by considering which components, events and interactions must be included, how far each aspect can be parameterised with available data, and the limits of our computational resources. It is essential to build on a sound understanding of the biological system and of the questions that matter to biologists and clinicians. Ideally this knowledge should come through close collaboration with empirical researchers throughout the model development process.

A general principle is that model complexity should match the complexity only of the phenomena of interest. We need not employ an off-lattice hybrid SABM if a simple cellular automaton with only a few basic update rules can demonstrate the same principle. Attempting to represent every component of a biological system is not only computationally impractical but also risks overfitting and hinders explainability. Simpler models have many merits. They are easier to falsify and have fewer sources of potential error. They reduce researcher degrees of freedom and curb the tweaking of parameters to support a pet hypothesis. They are more mathematically tractable and easier to analyse. Perhaps most importantly, a simple model has wider applicability and can be more readily generalised, adapted or extended to answer new questions. More complicated models should be preferred only if the biological system is especially well understood or if simpler models have been tested and shown to be inadequate.

Model design remains a challenge for even the most experienced researchers. One of the nine overarching themes in a recent review of key questions concerning the ecology and evolution of cancer \cite{dujon2021identifying} was that we do not yet know which mathematical and computational models are the most useful. In another recent survey of cancer adaptive therapy modelling \cite{west2023survey}, four of the eleven key open questions were related to identifying appropriate mathematical models. When it comes to SABMs, the main limitations are twofold. First, we typically lack sufficient data to design and parameterise SABMs of large tumours. Second, routinely simulating much more than a few million individual cells (corresponding to no more than half a cubic centimetre of tumour) is computationally impractical. To some extent, these problems have technological solutions. Multi-region sequencing, spatial multi-omics, digital pathology, and other modern methods are producing ever more detailed spatial tumour data. Accessible computing power continues to grow. But progress will also depend on developing smarter models.

Instead of drawing conclusions from a single SABM, we might do better to consider ensembles of models with diverse structures, algorithms, and underlying assumptions. Much as in hurricane forecasting \cite{hamill2012noaa}, we can be more confident when many models converge on the same prediction. Another important direction is to develop coarse-grained models that can simulate tumour evolution as accurately as cell-level SABMs but with much greater computational efficiency. Rather than cell division, death, mutation and dispersal rates, coarse-grained models depend on macroscopic parameters such as the arrival rate of consequential clones, clonal expansion speeds, and large-scale microenvironmental heterogeneity. A potential way forward is to combine mathematical analysis of the relevant stochastic processes to determine appropriate approximations, and machine learning methods to infer the parameter values. SABMs capable of accurately simulating the evolution of entire tumours could have wide-ranging applications, not least in patient-specific clinical forecasting.




\bibliographystyle{unsrt.bst}
\bibliography{ABM_9_Bib}

\appendix

\section*{Appendix: Agent-based modelling platforms}

\subsection*{Cell-based Chaste}
The cell-based version of Chaste \cite{mirams2013chaste} is a highly sophisticated, multiscale computational framework for modelling cell populations. Chaste permits both on- and off-lattice models and has built-in code for simulation of specific biological systems, such as cancer development within colonic crypts. Chaste has its own ODE and PDE solver, called SUNDIALS \cite{hindmarsh2005sundials}.

\subsection*{CompuCell3D}
CompuCell3D \cite{swat2012multi} is a general-purpose platform for implementing tissue development models, including the Glazier-Graner-Hogeweg (or cellular Potts) model that its developers pioneered. Its bespoke CC3D-Bionetsolver package solves ODEs and PDEs using a finite element method. CompuCell3D has been used in dozens of studies of cancer and morphogenesis.

\subsection*{Demon}
Demon \cite{nobledemon} specializes in simulating intratumour population genetics. Its multi-scale spatial structure makes it especially well suited to studying the evolution of glandular tumours. Demon can be configured to implement mathematically tractable models such as the Eden growth model, biassed voter model, spatial Moran process, and spatial branching processes. An automated computational workflow called warlock \cite{bak2023warlock} facilitates running demon simulations in parallel on a high-performance computing cluster.

\subsection*{HAL}
HAL \cite{westhal} is a generic and highly customisable platform comprised of modular components which allow for multiple grids to operate simultaneously, each performing different tasks. For example, one grid might handle cell-cell interactions while another implements oxygen diffusion. HAL has multiple ODE and PDE solvers to suit different modelling needs. It also includes several pre-built model templates.

\subsection*{J-SPACE}
J-SPACE \cite{angaroni2022j} is a modelling platform designed specifically for phylogenetic modelling. It simulates cancer evolution on a grid (or some other graph) and generates synthetic reads from next-generation sequencing platforms. A primary goal of J-SPACE is to help researchers assess the impact of incomplete data or experimental error on downstream bioinformatics pipelines.

\subsection*{PhysiCell}
PhysiCell \cite{ghaffarizadeh2018physicell} is a flexible framework that can implement physics-based off-lattice models of large numbers of cells in dynamic tissue microenvironments, with dynamic cell-cycle state tracking. PhysiCell uses a custom-built, open-source package for ODE and PDE solving, called BioFVM \cite{ghaffarizadeh2016biofvm}. Potential functions are used to describe cell-cell interactions including adhesion, repulsion, and cell-matrix interactions \cite{macklin2012patient}.

\subsection*{SMITH}
SMITH \cite{streck2023smith} implements a branching process with quasi-spatial constraints that separate the tumour into a proliferating shell and a static core. By simulating the dynamics of clones rather than individual cells, SMITH is able to simulate the evolution of a tumour up to a billion cells in only a few minutes on a standard desktop PC. This computational speed comes at the cost of the model's strong simplifying assumptions.

\subsection*{Morpheus}

Morpheus \cite{starruss2014morpheus} is a highly accessible open-source platform in which users can develop multi-scale, multicellular systems which couple ODEs, PDEs and cellular Potts models, with automatic scheduling. Rather than coding models manually, users can describe the model in biological and mathematical terms in Morpheus' GUI, and utilise provided tools for visualisation and parameter estimation. 

\end{document}